\documentclass[
 reprint,
superscriptaddress,
 amsmath,amssymb,
 aps,
]{revtex4-2}

\usepackage{graphicx}
\usepackage{dcolumn}
\usepackage{bm}
\usepackage[detect-all]{siunitx}
 \DeclareSIUnit\bar{bar}
\usepackage{lineno}
\usepackage{textgreek}
\usepackage{natbib}
\usepackage[english]{babel}
\newcommand{\supplementarysection}{%
  \setcounter{figure}{0}% Reset figure counter
  \let\oldthefigure\thefigure% Capture figure numbering scheme
  \renewcommand{\thefigure}{S\oldthefigure}% Prefix figure number with S

  \setcounter{table}{0}% Reset figure counter
  \let\oldthetable\thetable% Capture figure numbering scheme
  \renewcommand{\thetable}{S\oldthetable}% Prefix figure number with S
}

\begin{document}

\preprint{APS/123-QED}

\title{Coherent x-ray magnetic imaging with 5 nm resolution}

\author{Riccardo Battistelli}
\affiliation{Helmholtz-Zentrum Berlin, 14109 Berlin, Germany}
\affiliation{Experimental Physics V, Center for Electronic Correlations and Magnetism, University of Augsburg, 86159 Augsburg, Germany}
\author{Daniel Metternich}
\affiliation{Helmholtz-Zentrum Berlin, 14109 Berlin, Germany}
\affiliation{Experimental Physics V, Center for Electronic Correlations and Magnetism, University of Augsburg, 86159 Augsburg, Germany}
\author{Michael Schneider}
\affiliation{Max Born Institute for Nonlinear Optics and Short Pulse Spectroscopy, 12489 Berlin, Germany}
\author{Lisa-Marie Kern}
\affiliation{Max Born Institute for Nonlinear Optics and Short Pulse Spectroscopy, 12489 Berlin, Germany}
\author{Kai Litzius}
\affiliation{Experimental Physics V, Center for Electronic Correlations and Magnetism, University of Augsburg, 86159 Augsburg, Germany}
\author{Josefin Fuchs}
\affiliation{Max Born Institute for Nonlinear Optics and Short Pulse Spectroscopy, 12489 Berlin, Germany}
\author{Christopher Klose}
\affiliation{Max Born Institute for Nonlinear Optics and Short Pulse Spectroscopy, 12489 Berlin, Germany}
\author{Kathinka Gerlinger}
\affiliation{Max Born Institute for Nonlinear Optics and Short Pulse Spectroscopy, 12489 Berlin, Germany}
\author{Kai Bagschik}
\affiliation{Deutsches Elektronen-Synchrotron (DESY), 22607 Hamburg, Germany}
\author{Christian M. G\"unther}
\affiliation{Technische Universit\"at Berlin, Zentraleinrichtung Elektronenmikroskopie (ZELMI), 10623 Berlin, Germany}
\author{Dieter Engel}
\affiliation{Max Born Institute for Nonlinear Optics and Short Pulse Spectroscopy, 12489 Berlin, Germany}
\author{Claus Ropers}
\affiliation{Max Planck Institute for Multidisciplinary Sciences, 37077 G\"ottingen, Germany}
\affiliation{4th Physical Institute, University of G\"ottingen, G\"ottingen, Germany}
\author{Stefan Eisebitt}
\affiliation{Max Born Institute for Nonlinear Optics and Short Pulse Spectroscopy, 12489 Berlin, Germany}
\affiliation{Technische Universit\"at Berlin, Institut f\"ur Optik und Atomare Physik, 10623 Berlin, Germany}
\author{Bastian Pfau}
\affiliation{Max Born Institute for Nonlinear Optics and Short Pulse Spectroscopy, 12489 Berlin, Germany}
\author{Felix B\"uttner}
\email{felix.buettner@uni-a.de}
\affiliation{Helmholtz-Zentrum Berlin, 14109 Berlin, Germany}
\affiliation{Experimental Physics V, Center for Electronic Correlations and Magnetism, University of Augsburg, 86159 Augsburg, Germany}
\author{Sergey Zayko}
\affiliation{Max Planck Institute for Multidisciplinary Sciences, 37077 G\"ottingen, Germany}
\affiliation{4th Physical Institute, University of G\"ottingen, G\"ottingen, Germany}

\date{September 18, 2023}

\begin{abstract}
Soft x-ray microscopy plays an important role in modern spintronics.
However, the achievable resolution of most x-ray magnetic imaging experiments is above \SI{10}{\nm}, limiting access to fundamental and technologically relevant length scales.
Here, we demonstrate x-ray magnetic microscopy with \SI{5}{\nm} resolution by combining holography-assisted coherent diffractive imaging with heterodyne amplification of the weak magnetic signal.
The gain in resolution and contrast allows direct access to key magnetic properties, including domain wall profiles and the position of pinning sites.
The ability to detect and map such properties with photons opens new horizons for element-specific, time-resolved, and \textit{in-operando} research on magnetic materials and beyond.
\end{abstract}

\maketitle

\section{Introduction}
Nanometer-scale textures are ubiquitous in nature and can emerge due to competing interactions~\cite{seul_domain_1995}, topological constraints or pinning at defects.
In the realm of magnetism, prominent examples are domain walls, vortices, and skyrmions in two dimensions, as well as even richer topological configurations in three dimensions~\cite{gobel_beyond_2021}.
These objects are subject to intense research---both because of their exciting physics and their potential use in unconventional, low-energy, high-performance data storage, computing and sensing technologies~\cite{vakili_skyrmionicscomputing_2021-1}.

Progress in this field relies on high resolution imaging techniques such as scanning probe techniques, Lorentz microscopy and magneto-optical methods~\cite{Reeve2020}.
In this context, soft x-ray imaging methods stand out as \textit{in-operando} techniques because they allow to combine nanometer spatial resolution, pico- to femtosecond temporal resolution, element-specific contrast, and the ability to study patterned devices during the application of a variety of stimuli~\cite{10.1063/1.4942776,van_waeyenberge_magnetic_2006,baumgartner_spatially_2017, donnelly_three-dimensional_2017,caretta_fast_2018,https://doi.org/10.1002/adma.201807683,klose_coherent_2023}.
X-ray imaging has, for example, enabled the direct observation of magnetic vortex core switching~\cite{van_waeyenberge_magnetic_2006}, spin-orbit torque switching of a magnetic random access memory layer~\cite{baumgartner_spatially_2017}, the 3D spin configuration around Bloch points~\cite{donnelly_three-dimensional_2017} and, owing to element-specific contrast, skyrmions in a compensated ferrimagnet~\cite{caretta_fast_2018}.
However, despite advancements in this field, achieving a resolution better than \SI{20}{\nm} remains challenging, and only exceptional experiments have reached 10 nm resolution in x-ray magnetic imaging~\cite{10.1063/1.4942776,https://doi.org/10.1002/adma.201807683}.
Consequently, photon-based microscopy has been limited in its ability to resolve some of the most fundamental and technologically relevant properties of magnetic materials such as nanometer-scale domain walls and their relation with magnetic inhomogeneities.

In this study, we advance photon-based magnetic imaging to a resolution of \SI{5}{\nm}, already on a third-generation synchrotron-radiation source. Specifically, we employ holography-assisted phase retrieval (HAPRE)~\cite{kfir_nanoscale_2017} using an approach based on two ingredients: first, waveguiding reference holes with a modulated exit wave~\cite{zayko_coherent_2015,malm_reference_2022} to amplify the weak magnetic scattering at large scattering angles~\cite{zayko_ultrafast_2021, klose_photon_2022} and second, a combination of Fourier transform holography (FTH)~\cite{eisebitt_lensless_2004,flewett_holographically_2012} and iterative phase retrieval algorithms~\cite{luke_relaxed_2005,fienup_phase_1982}.
The method allows us to resolve magnetic domain walls and magnetic pinning points in a Pt/Co-type multilayer. The results provide a qualitatively new level of understanding of the local interactions at play in this otherwise well-studied class of materials.

\section{Materials and Methods}
\begin{figure}[h]
\includegraphics[width=\linewidth]{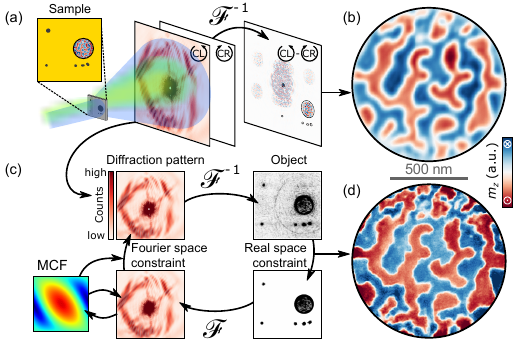}
\caption{The HAPRE workflow.
(a) A coherent x-ray beam of circularly left-handed (CL) and right-handed (CR) polarization is focused on the sample and the resulting diffraction patterns are recorded with a CCD camera.
(b) Magnetic state (as the difference between the FTH-reconstructed CL and CR exit waves).
(c) Sketch of the iterative phase retrieval process. 
(d) Phase-retrieved magnetic image, obtained from the reconstruction of both polarizations.
}
\label{fig:fig1}
\end{figure}

The sample is a $\mathrm{Pt}(3)|[\mathrm{Pt}(2)|\mathrm{Co}(0.8)|\mathrm{Cu}(0.5)]_{15}|\mathrm{Pt}(2)$ (thicknesses in \SI{}{\nm}) multilayer grown via direct current magnetron sputtering on a SiN membrane.
The material belongs to the class of chiral magnetic multilayers, which are of large interest in the field of spintronics~\cite{gobel_beyond_2021,vakili_skyrmionicscomputing_2021-1} since in such materials the combination of perpendicular magnetic anisotropy and Dzyaloshinskii-Moriya interaction (DMI) leads to the emergence of nanometer-scale magnetic textures.
Imaging was performed in a FTH geometry~\cite{eisebitt_lensless_2004}~[Fig.~\ref{fig:fig1}(a)] at the Co L\textsubscript{3}-edge (wavelength \SI{1.59}{nm}) using x-ray magnetic circular dichroism contrast.
To realize FTH, we used a monolithically integrated opaque Cr/Au mask with a \SI{1}{\um} diameter object hole and reference holes with diameters of \SI{40}{}\text{--}\SI{60}{\nm} arranged in a pattern optimized for HAPRE~\cite{zayko_ultrafast_2021}.

To account for the dynamic range of the diffraction pattern, we recorded and merged data with increasing beam-stop sizes.
The scattering data was offset corrected, normalized and projected from the flat detector plane onto the Ewald sphere to remove spherical aberrations. See Appendix \ref{appendix:Supplement_1} for details.

The FTH mask design allows us to obtain a magnetic image (“reconstruction”) from a single Fourier transform~[Fig.~\ref{fig:fig1}(b)] as a convolution between the object and reference hole exit waves.
As a consequence, the resolution and contrast of this reconstruction are limited by the reference hole size and shape~\cite{eisebitt_lensless_2004}.
However, this relatively low-resolution image is sufficient to accurately determine the support mask needed for iterative phase-retrieval algorithms~[Fig.~\ref{fig:fig1}(c)], which removes the FTH limitation.
We first reconstructed the left-circularly polarized exit wave $\phi_{CL}$ with a combination of the relaxed averaged alternating reflections (RAAR) algorithm (700 iteration steps)~\cite{luke_relaxed_2005}, and error-reduction (ER) algorithm (50 iteration steps)~\cite{fienup_phase_1982}.
Then, the right-circularly polarized exit wave $\phi_{CR}$ was reconstructed employing only the ER algorithm (50 iteration steps) using the phase of $\phi_{CL}$ as a starting guess.
Finally, our reconstruction routine accounts for partial coherence (e.g., due to low intrinsic spatial coherence of the source, vibrations in the beam path, and contributions of the higher harmonics, as known at the beamline~\cite{bagschik_direct_2020}) by incorporating a Richardson-Lucy deconvolution algorithm~\cite{clark_high-resolution_2012}.

The local, thickness-integrated out-of-plane magnetization was extracted as $m_z \propto \log \left( \lvert \phi_{CL}/\phi_{CR} \lvert \right) $ and focused numerically~\cite{malm_reference_2022}. The result is a purely magnetic image independent of topographic variations and linearly proportional to the magnetization~[Fig.~\ref{fig:fig1}(d)]~\cite{kfir_nanoscale_2017}.
Note that absolute values for $m_z$ can in principle be obtained, but this information was practically inaccessible in our experiment due to higher harmonic contributions at low spatial frequencies and fluctuations of the illumination intensity~\cite{bagschik_direct_2020}.
It is expected that these challenges will be addressed at the next generation of x-ray sources.

\section{Results and Discussion}

\begin{figure}[h]
\includegraphics[width=\linewidth]{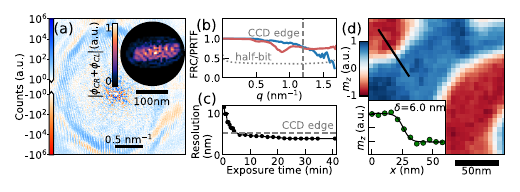}
\caption{Demonstration of magnetic imaging with \SI{5}{nm} resolution.
(a) Magnetic diffraction pattern ($I_\text{diff}$) corresponding to Fig.~\ref{fig:fig1}(d). Inset: Reconstructed exit wave of a reference hole exhibiting strong modulations.
(b) Fourier ring correlation (FRC, blue line) and phase-retrieval transfer function (PRTF, red line) of the magnetic reconstruction obtained from (a).
(c) Spatial resolution versus total exposure time, computed with FRC.
(d) Section of Fig.~\ref{fig:fig1}(d). Inset: Magnetization profile $m_z(x)$ (dots) measured along the black line in the main panel. The line is a fit to $m_z(x)=\tanh((x-x_0)/\delta)$.
(a.u., arbitrary units).
}
\label{fig:fig2}
\end{figure}

In our masked-based thin-film sample geometry, the dominant magnetic signal in the diffraction pattern is the heterodyne product $\Psi_c\Psi_m$ between the charge scattering of the mask ($\Psi_c\propto\mathcal{F}[T_\text{mask}(\mathbf{r})]$, where $T_\text{mask}(\mathbf{r})$ is the transmission function of the mask including the reference holes) and the magnetic scattering of the spin texture ($\Psi_m\propto\pm\mathcal{F}[ T_\text{mask}(\mathbf{r})m_z(\mathbf{r})]$, where the sign alternates between CL and CR polarization)~\cite{klose_photon_2022,kfir_nanoscale_2017}.
The magnetic signal can be isolated in the difference diffraction pattern $I_\text{diff}(\mathbf{q})=I_{CR} - I_{CL} \propto \mathrm{Re}(\Psi_c\Psi_m)$, as depicted in Fig.~\ref{fig:fig2}(a).
Importantly, the signal at high photon momentum transfer $\mathbf{q}$, where the information from the smallest magnetic features is encoded, is extremely weak.
Thus, high-resolution magnetic imaging relies on charge scattering up to the highest $\mathbf{q}$ to amplify weak magnetic scattering above the detector noise level and above the minimum of one photon per speckle.

To obtain charge scattering at high $\mathbf{q}$, we exploit the 3D geometry of the reference holes to realize strong modulations in their exit waves based on waveguide mode beating~\cite{zayko_coherent_2015,malm_reference_2022,kfir_nanoscale_2017}.
Specifically, we produced three modified reference holes with irregular shapes, which have exit waves with sub-\SI{10}{nm} internal modulations as revealed by real-space reconstructions [see inset Fig.~\ref{fig:fig2}(a)].
The result is a strong difference signal up to the edges of the detector~[Fig.~\ref{fig:fig2}(a)], corresponding to a resolution of the retrieved images of \SI{5.3}{\nm}. The signal remains strong even to the corners of the detector, demonstrating that a resolution of \SI{3.7}{\nm} is immediately in reach with a larger detector. Fourier ring correlation and phase-retrieval transfer function measurements verify these values~[Fig.~\ref{fig:fig2}(b)] (see Appendix \ref{appendix:Supplement_1}).
An analysis of the resolution as a function of the total exposure time reveals that 5 minutes were sufficient to reach this result, although longer integration times yields higher signal-to-noise ratio (see Appendix \ref{appendix:Supplement_1}).
Figure~\ref{fig:fig2}(d) displays an image and a line scan of a magnetic domain wall, from which we measure a domain wall width parameter $\delta$ of \SI{6}{\nm} (full domain wall width is $\pi\delta$).
This is close to the domain wall width parameter expected in our material, $\delta_0=\SI{5.1}{nm}$, as estimated with micromagnetic simulations, thus confirming our resolution estimate in real space (see Appendix \ref{appendix:Supplement_1}).
The achieved resolution is the highest of any photon-based magnetic imaging technique reported so far.

\begin{figure}[h]
\includegraphics[width=\linewidth]{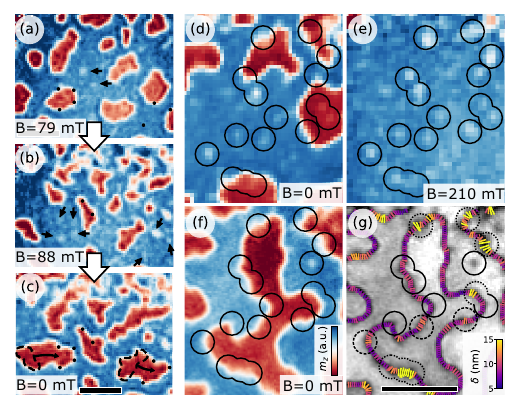}
\caption{High-resolution imaging of magnetic pinning. (a-f) Domain patterns. (g) Measured domain wall widths of the domain state in (f). Annotations highlight the locations of magnetic pinning points. Details, see text. Scalebar in (c) refers to (a-c), scalebar in (g) refers to (d-g). Scalebars are \SI{300}{\nm}.}
\label{fig:fig3}
\end{figure}
The enhanced contrast and resolution of HAPRE imaging not only improves the visibility of known magnetic features, but also provides access to properties previously invisible in photon-based magnetic imaging experiments.
One of these properties is magnetic pinning – the ubiquitous phenomena that magnetic textures get pinned at certain locations in the material.
A long-standing challenge in materials science is to detect the location and the underlying mechanism of pinning~\cite{klose_coherent_2023,gruber_skyrmion_2022}. HAPRE can address both aspects, as we demonstrate next.

As an example, we study magnetic pinning in our multilayer after fs laser exposure.
This choice is motivated by the wide interest in laser-induced ultrafast dynamics in chiral materials and by the importance of understanding microscopic laser-induced damage and pinning in this context~\cite{stark_controlling_2015}.
In a first experiment we imaged the domain state with a pixel size of \SI{10}{\nm}.
Figures~\ref{fig:fig3}(a-c) display a section of the experimental field of view after various levels of exposure to a \SI{1030}{\nm} fiber laser with \SI{250}{\fs} pulses.
Moderate laser exposure (10 pulses, at \SI{79}{\milli\tesla} magnetic field) led to the anticipated nucleation of isolated domains~[Fig.~\ref{fig:fig3}(a)].
However, more intense laser exposure (100 pulses with fluence increased by a factor 1.3, at \SI{88}{\milli\tesla}) additionally resulted in large damaged areas characterized by smaller domains and reduced magnetic contrast [top of Fig.~\ref{fig:fig3}(b,c)].
The subsequent removal of the magnetic field led to the expansion of magnetic domains~[Fig.~\ref{fig:fig3}(c)].
Throughout this sequence of magnetic states, the presence of pinning sites can be inferred by identifying locations where domain walls repeatedly come to rest or form cusps, as indicated by black dots.
Moreover, we observe indications of direct visibility of the pinning sites even in the absence of a domain wall, which appear as roundish objects exhibiting reduced magnetic contrast, as highlighted by black arrows.
However, their similarity with noise makes precise verification challenging in these images.

The signal-to-noise ratio can be increased with longer exposure times (see Appendix \ref{appendix:Supplement_1}), as shown in Fig.~\ref{fig:fig3}(d,e) for a different field of view on the same sample as before.
This allows to robustly detect the low-contrast features.
Yet another level of insight is reached by enhancing the resolution to \SI{5}{\nm}, which enables imaging of domain walls and measurement of their local width \textdelta\ [Fig.~\ref{fig:fig3}(f,g)].
To explore the relation between low-contrast features and domain wall pinning, we imaged the sample at remanence~[Fig.~\ref{fig:fig3}(d)], following a partial demagnetization cycle which promoted domain rearrangement~[Fig.~\ref{fig:fig3}(f)] and under an external field of \SI{210}{\milli\tesla}~[Fig.~\ref{fig:fig3}(e)].
The low-contrast features consistently appeared in the same positions (see circles in Fig.~\ref{fig:fig3}(d-e)) and were not annihilated by the external field, indicating a strong connection to specific locations.
Furthermore, the images confirm an attraction between these locations and domain walls (see Fig.~\ref{fig:fig3}(f) for examples) and reveal a complex relationship between the two.
In some cases the domain wall width \textdelta\ at a defect is comparable to defect-free regions of the sample (solid circles), while in others the domain walls are larger (dashed circles), with values exceeding \SI{15}{\nm} (dotted circles).

These findings indicate that the low-contrast features are found at pinning points of the material with altered magnetic or structural properties.
The direct observation of small, low magnetization spots at pinning points was already reported in CrBr\textsubscript{3} bilayers using NV center scanning magnetometry, although their magnetization pattern could not be spatially resolved~\cite{sun_magnetic_2021}.

\begin{figure}[h]
\includegraphics[width=\linewidth]{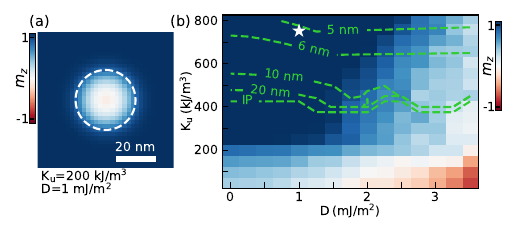}
\caption{Micromagnetic simulation of the visibility of pinning points.
(a) An example magnetic state in a defect of reduced anisotropy. The dashed line indicates extent of the \SI{30}{\nm}-diameter defect.
(b) Phase space of $m_z$ found at the center of the defect as a function of $K_\text{u}$ and $D$. The white star marks the parameter values of the defect-free region of material.
Green contour lines mark the dependence of \textdelta\ on $K_\text{u}$ and $D$, as determined from simulations of a uniform magnetic material.}
\label{fig:fig4}
\end{figure}
In multilayer materials the interface-related magnetic parameters such as the DMI ($D$) and uniaxial anisotropy ($K_\text{u}$) can easily be affected by growth conditions, roughness of the substrate and post-growth treatment~\cite{stark_controlling_2015,bertero_perpendicular_1994,PhysRevB.105.064429}.
Additionally, local variations of these parameters explain the presence of attractive pinning points in the material, as an increase in $D$ or a reduction in $K_\text{u}$ would lower the domain wall energy density $\sigma$. (In first approximation, $\sigma = 4\sqrt{A K_{\text{eff}}}-\pi \lvert D \rvert$~\cite{PhysRevB.105.064429}, where $A$ is the exchange stiffness, $K_\text{eff}=K_\text{u}-\mu_0 M_\text{s}^2/2$ and $M_\text{s}$ is the saturation magnetization.)
Furthermore, the width of domain walls \textdelta\ would increase with lower $K_\text{u}$ while being independent of $D$, as predicted by theoretical models ($\delta \approx \sqrt{A/K_{\text{eff}}}$ ) and confirmed with micromagnetic simulations~[Fig.~\ref{fig:fig4}(b)].
Therefore, localized independent variations of $K_\text{u}$ and $D$ are both plausible and explain the variety of pinned domain wall widths observed experimentally in Fig.~\ref{fig:fig3}(f): while high-DMI defects attract domain walls without affecting the domain wall width, low-anisotropy defects also increase \textdelta.

To understand how these defects would also be directly visible as points with lower magnetic contrast, we performed micromagnetic simulations using the Mumax3 solver~\cite{Vansteenkiste2014}.
The bulk of the material was modeled with magnetic parameters derived from experimental data and literature, while the $K_\text{u}$ and $D$ were altered inside a circular defect of \SI{30}{\nm} in diameter.
The simulation was started from a random magnetization and relaxed under an external field of \SI{210}{\milli\tesla}.
After relaxation, the major part of the sample is saturated (with $m_z=1$), while for certain $K_\text{u}$ and $D$ the defect stabilizes spin textures locally lowering the net magnetization $m_z$ at its center, as observed in Fig.~\ref{fig:fig4}(a).
Systematic exploration of the anisotropy-DMI phase space~[Fig.~\ref{fig:fig4}(b)] reveals that a local reduction of $m_z$ can robustly be achieved if within the defect either $K_\text{u}$ is lowered or DMI is increased.
The same trends also lower the domain wall energy, which thus explains why low-contrast features are pinning sites for domain walls.

\section{Conclusions}
We have demonstrated x-ray magnetic imaging with a record \SI{5.3}{\nm} resolution using holography-assisted phase retrieval. A resolution of \SI{3.7}{\nm} is immediately in reach with larger detectors.
Key to this ability is a strong charge scattering at high $\mathbf{q}$, achievable with a reference beam modulated by waveguide mode beating.
Our high-resolution images resolve the structure of magnetic domain walls, make magnetic pinning directly visible, and allow to investigate the underlying pinning mechanisms. These capabilities lift photon-based magnetic microscopy to a qualitatively new level.
The concept of heterodyne amplification of weak signals can readily be extended to other, maskless coherent imaging approaches and is fully compatible with time-resolved coherent imaging~\cite{klose_coherent_2023,zayko_ultrafast_2021}.
Most importantly, the technique fully leverages the emergence of high-coherence x-ray sources around the world, where sub-wavelength spatial resolution (as demonstrated in the extreme UV regime~\cite{zayko_ultrafast_2021}) comes in reach, thereby finally extending x-ray magnetic imaging to the fundamental length scales as defined by the exchange interaction.

\supplementarysection
\begin{appendix}
\section{Extended materials and methods}\label{appendix:Supplement_1}

This appendix describes the experimental and post-processing procedures used to obtain the results presented in the main text.

\subsection{Sample fabrication}
The magnetic multilayer presented in this paper has the following structure: $\mathrm{Pt}(3)|[\mathrm{Pt}(2)|\mathrm{Co}(0.8)|\mathrm{Cu}(0.5)]_{15}|\mathrm{Pt}(2)$ (thicknesses in \SI{}{\nm}). The sample was prepared at room temperature via direct current magnetron sputtering from elemental targets on 100-nm-thick silicon-nitride membranes. The sputter process was carried out in a vacuum chamber (base pressure lower than \SI{5e-6}{\milli\bar}) using an Ar working pressure of \SI{1.2e-3}{\milli\bar}.
The sample was covered with a [Cr(\SI{5}{\nm})/Au(\SI{50}{\nm})]\textsubscript{30} x-ray opaque film grown via thermal evaporation on the side of the SiN membrane opposite to the magnetic film. 
To enable Fourier transform holography, a circular aperture of \SI{1}{\um} in diameter and additional pinholes (\SI{40}{}\text{--}\SI{60}{\nm} in diameter) were carved in the Cr/Au film, acting as the field of view (FOV) and as the holographic reference apertures necessary for the technique, respectively. Additional auxiliary apertures were added to the design, to amplify the scattered signal to larger angles \cite{kfir_nanoscale_2017}, see Fig.~\ref{fig:figS1}(g,f) in particular. The apertures were carved on the mask using focused ion beam milling. Ga\textsuperscript{+} ions at \SI{8}{} and \SI{30}{\kilo\volt} were used for the FOV and reference apertures, respectively.

\subsection{Imaging setup}
The x-ray imaging experiments were carried at the P04 beamline of the PETRA III synchrotron-radiation facility of Hamburg at the L\textsubscript{3}-edge of cobalt (wavelength \SI{1.59}{\nm}). Magnetic contrast was achieved exploiting the XMCD effect by recording transmission diffraction patterns using light with left and right circular polarization. The sample was placed roughly at the focus of the beamline and a CCD camera was placed downstream to record the resulting diffraction pattern. Details on the camera employed, its distance from the sample and the size of the pixels in the reconstructions are listed in Table \ref{tab:images_exp_details} for every image displayed in the main text.

To ensure correct sampling of the entire diffraction patterns, we recorded them multiple times using two beam-stops of different sizes, allowing us to protect the camera from overexposure and increase the exposure time to obtain higher signal to noise ratio.
Table \ref{tab:images_exp_times} contains relevant information on the beam-stops used for the reconstruction with \SI{5}{\nm} resolution and the number and exposure times of the frames recorded with them.

The results were merged into composite images with maximized dynamic range, ensuring the correct sampling of small and large scattering angle signal.
The recorded diffraction pattern were then projected onto the Ewald sphere to account for aberration due to the flat detector \cite{schaffert_high-resolution_2013} and centered. Finally, the left and right polarization diffraction patterns were normalized to each other in order to account for different intensities of opposite polarized illumination and offset corrected in order to eliminate the constant contribution of the camera readout.

The out-of-plane magnetic field at the sample was applied using an electromagnet.
For the measurements after laser exposure shown in Figure~3, a \SI{1030}{\nm} wavelength fiber laser was focused on the sample at normal incidence. Laser pulses were \SI{250}{\fs} long.
For details on the laser setup, see Gerlinger et al. \cite{gerlinger_application_2021}.

\subsection{Holography-assisted phase retrieval}
The images were first reconstructed via a simple Fourier transform to obtain an immediate real-time feedback.
To increase resolution and contrast, we then applied iterative phase retrieval algorithms to recover the missing phase lost upon recording.
To this end, we first defined the support function $s(\mathbf{r})$ for the phase retrieval algorithm manually, using the obtained FTH reconstruction as a reference [Fig.~\ref{fig:figS2}]. This allowed us to define a precise and tight support, facilitating rapid convergence to an appropriate solution.
The starting guess $\Psi_i$ for the complex diffraction pattern was estimated by multiplying the support function by a constant $c$, and then performing its Fourier transform.
\begin{equation}\label{eq:0}
\begin{aligned}
\Psi_i(\mathbf{q}) =  \mathcal{F} \left[ c \cdot s(\mathbf{r}) \right].
\end{aligned}
\end{equation}

The phase of $\Psi_i(\mathbf{q})$ was used to estimate the phase lost upon recording on the entirety of diffraction pattern, while its square modulus was used to estimate the diffraction pattern intensity at pixels where signal was not recorded reliably (e.g. damaged, unreliable or covered pixels of the CCD camera). In particular, the first Airy disk of the diffraction pattern includes relevant spurious contributions from higher harmonics of the fundamental \SI{1.59}{\nm} wavelength beam (typically unavoidable at synchrotron facilities), posing questions of compatibility with the rest of the recorded data. We recorded the first Airy disk information for reconstructions depicted in Figure~3(c,d), but left it covered with a beam-stop for the reconstruction in Figure~3(a,b,e,f). In this last case, we used the square modulus of $\Psi_i(\mathbf{q})$ to initially estimate the diffraction pattern intensity at the missing pixels, and then updated these values during the iterative phase retrieval by substituting the missing information with reconstructed amplitudes \cite{latychevskaia_iterative_2018}.
We found this approach to generally guarantee appropriate convergence to a valid solution.

We then reconstructed the phase of our recorded diffraction patterns using a two-step protocol.
In the first step, we used a combination of the relaxed averaged alternating reflections \cite{luke_relaxed_2005} (RAAR) and error reduction \cite{fienup_phase_1982} (ER) phase retrieval algorithms. Precise alignment between the two opposite polarized images was ensured by adopting a workflow suitable to dichroic imaging \cite{kfir_nanoscale_2017, zayko_ultrafast_2021}. First, we retrieved the phase of the left handed circular polarization image using 700 RAAR iterations (with the relaxation parameter \textbeta\ decreasing gradually from 1 to 0.5).
Then, we refined the reconstruction of the left-handed circular polarization image with an additional 50 ER iterations. The retrieved left-handed phase was then overlaid to the right handed diffraction pattern and the corresponding right handed phase was retrieved with just 50 ER iterations. This rapid convergence to a solution can be verified by observing the trend of the error metric (as defined by Clark et al. \cite{clark_high-resolution_2012}) displayed in Figure~\ref{fig:figS3}, which for the right handed reconstruction converges to a constant value after a few iteration steps.
In the second step of our protocol, the phase retrieval process was also repeated a second time using a relaxed magnitude constraint \cite{clark_high-resolution_2012} to compensate for artifacts arising from partially incoherent illumination and/or sample drift, such as blurring of the recorded diffraction pattern or the presence of incoherent background.
The partially coherent intensity pattern recorded in the experiment can be described as:

\begin{equation}\label{eq:1}
\begin{aligned}
I_{\text{pc}}(\mathbf{q}) = I(\mathbf{q}) \otimes \hat{\gamma}(\mathbf{q})
\end{aligned}
\end{equation}
where $I(\mathbf{q})$ is the hologram that would be recorded in case of perfectly coherent light and no drift and $\hat{\gamma}(\mathbf{q})$ is the Fourier transform of the normalized mutual coherent function of the illumination. By estimating $\hat{\gamma}(\mathbf{q})$, the coherent diffraction pattern $I(\mathbf{q})$ can be extracted from the experimentally recorded data and used for the phase retrieval algorithm, improving the stability of the algorithm and convergence to a better solution.
While the main phase retrieval algorithm was running, we recursively estimated $\hat{\gamma}(\mathbf{q})$ with a subroutine implementing the Richardson-Lucy (RL) deconvolution algorithm. $\hat{\gamma}(\mathbf{q})$ was updated every 20 iteration steps of the main phase retrieval algorithm, with the RL subroutine running for 50 iterations each time. In agreement with literature \cite{clark_high-resolution_2012}, we found that the phase retrieval algorithm adapted to partially coherent data generally outperforms algorithms assuming full coherence, both in terms of the error function and on the quality of reconstructed images [Fig.~\ref{fig:figS3}]. A detailed explanation of the Richardson-Lucy deconvolution algorithm is offered by Clark et al. \cite{clark_high-resolution_2012}.

\begin{figure}[h]
\centering
\includegraphics[width=12.9cm]{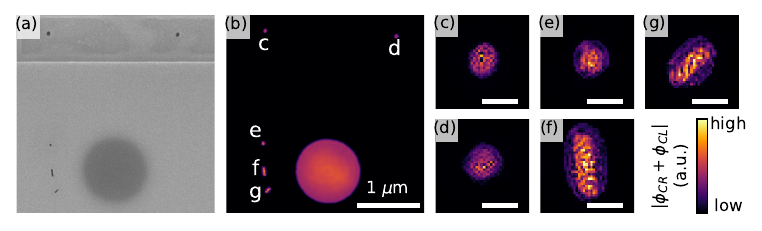}
\caption{Reference aperture exit wave modulation.
(a) Scanning electron micrograph of the holographic masks. (b) Respective transmission image of the samples as reconstructed by the phase retrieval algorithm. (c-g) Insets showing close-ups of the reconstructed reference hole exit waves. Scalebars in (c-f) are~\SI{100}{\nm}. (a.u., arbitrary units)}
\label{fig:figS1}
\end{figure}

\begin{figure}[h]
\centering
\includegraphics[width=8.79 cm]{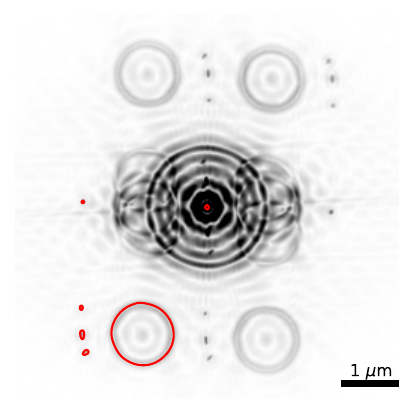}
\caption{FTH reconstruction (Patterson map) of the sample shown in Figure~3(a) (unpolarized image obtained by summing up the two opposite polarization images) and manually defined support function (delimited by the red lines).}
\label{fig:figS2}
\end{figure}

\begin{figure}[h]
\centering
\includegraphics[width=13 cm]{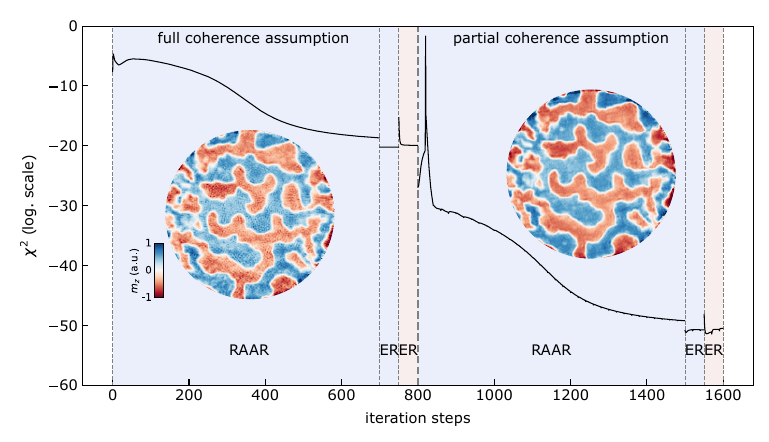}
\caption{Error metric as a function of the iteration number for the entire reconstruction process of the reconstruction displayed in Figure~3(a). The blue/red light background correspond to iteration steps reconstructing the left/right handed circular polarizations, respectively. The insets show the results of the first and second phase of our retrieval algorithm.}
\label{fig:figS3}
\end{figure}

Finally, the magnetic-contrast reconstruction was computed as:
\begin{equation}\label{eq:2}
\begin{aligned}
m_z \propto \log{\left(  \lvert \phi_{CL}/\phi_{CR}  \lvert \right)} 
\end{aligned}
\end{equation}
where $\phi_{CL}$ and $\phi_{CR}$ are the left and right circularly polarized reconstructions, respectively. The unpolarized transmission map showed in inset Fig.~\ref{fig:fig2}(a) and Fig.~\ref{fig:figS1} was obtained summing the left and right circularly polarized images.
The entire phase retrieval process takes approximately five minutes using a NVIDIA GeForce RTX 2070 SUPER GPU for the 2048\texttimes2048 images, while taking under two minutes for the 1340\texttimes1300 images.

\subsection{Domain wall measurement}
\begin{figure}[h]
\centering
\includegraphics[width= 8.79 cm]{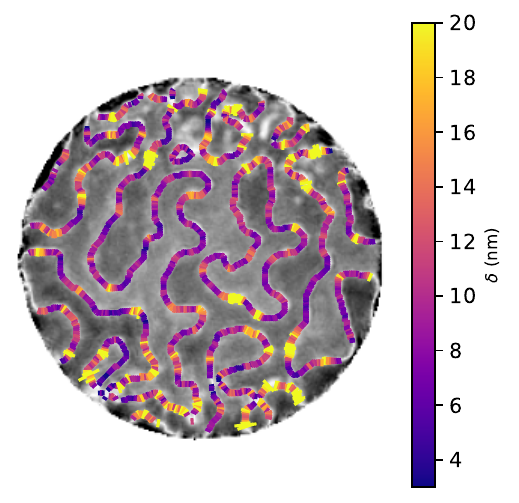}
\caption{Domain walls detected in the reconstructed image. The colored bars represent the detected domain walls. Each line is centered at the domain wall position, is perpendicular to the domain wall and has a length equal to the domain wall width \textdelta.}
\label{fig:figS4}
\end{figure}

Domain wall profiles were computed in two steps: (i) automatically detecting the domain walls out of focus to avoid bias and (ii) interpolating the data and capturing magnetization values perpendicular to the domain wall. The profiles were then fitted using the function $m(x)=\tanh(x/\delta)$, from which we obtained the domain wall width \textdelta. Figure~\ref{fig:figS4} displays the entire field of view for the reconstruction shown in Figure~3(f) with the detected and measured domain walls.

\subsection{Determination of resolution}
The resolution was determined using two separate methods: Fourier ring correlation (FRC) \cite{van_heel_fourier_2005} and the phase retrieval transfer function (PRTF) as defined by Marchesini et al. \cite{marchesini_phase_2005}.
FRC measurements were performed on two independent reconstructions, generated by dividing the data into two sets and performing holography aided phase retrieval reconstructions on both groups separately, with the support $s(\mathbf{r})$ as the only input in common.
To ensure that the measurements would be only sensitive to the magnetic signal, the FRC was computed on the Fourier transform of the magnetic reconstructions. Additionally, the magnetic reconstructions were multiplied by a function equal to one inside the field of view aperture but gradually decaying to zero at its edge to exclude spurious contributions from the sharp edges of the aperture. The resolution of the reconstructions was obtained from the computed FRC measurements using the half-bit threshold criterion \cite{van_heel_fourier_2005}.
The PRTF was computed for each polarization for 100 independent reconstructions. The reconstructions were started from random initial guesses for all CCD pixels with the exclusion of the pixels covered by the beam-stop, for which we always provided the same initial amplitude and phases in order to easen the alignment of each reconstruction.

\subsection{Time and efficiency considerations}

\begin{figure}[h]
\centering
\includegraphics[width= 13 cm]{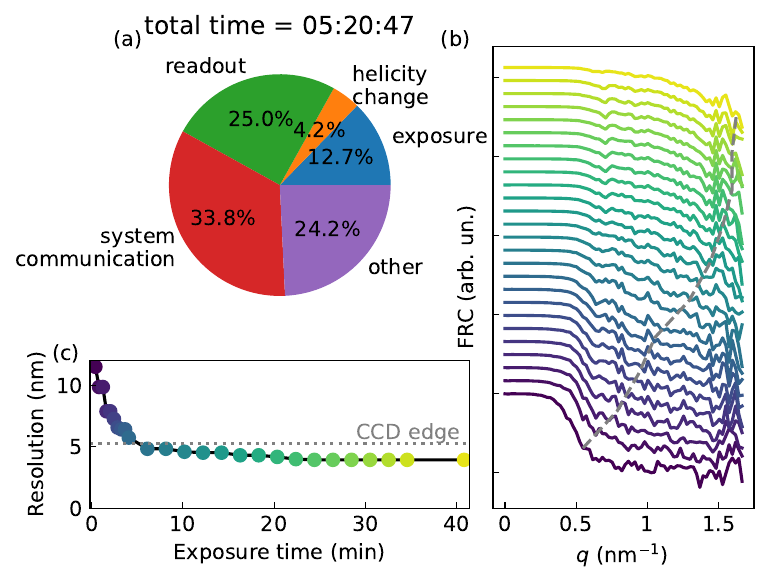}
\caption{Analysis of the experiment efficiency. (a) Time consumed for the acquisition of the image displayed in Figure~3(a), divided by category. (b) FRC curves of the reconstructed images at increasing exposure times. An offset has been added for clarity. The gray dashed line is a guide to the eye indicating the points where the FRC curves cross the half-bit thresholds. (c) Resolution of the reconstructions as a function of exposure time, taken from (b).}
\label{fig:figS5}
\end{figure}

\begin{figure}[h]
\centering
\includegraphics[width= 13 cm]{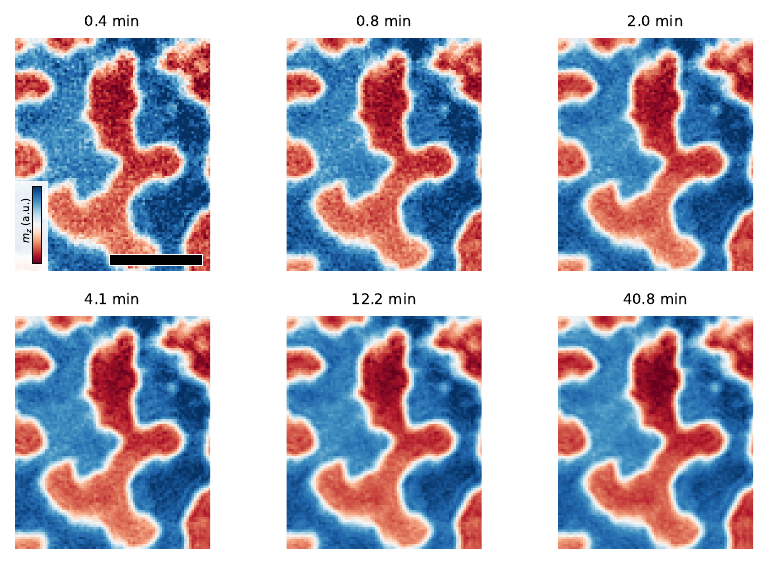}
\caption{Analysis of the experiment efficiency. Section of the field of view showed in Figure 3(f) using different exposure times. Scalebar is \SI{300}{\nm}.}
\label{fig:figS6}
\end{figure}

Figure~\ref{fig:figS5}(a) displays an analysis of how time was spent acquiring the image presented in Figure~3 of the main text.
While a total of 5 hours and 21 minutes were taken to complete the image recording, the total exposure time used only consisted of around 41 minutes.
Most of the time (3 hours and 9 minutes) was taken by camera readout and pauses due to inefficient communication protocols between the software used and the camera itself. As these issues have now been solved, we estimate the same experiment could be replicated in less than 2 hours and 40 minutes. 
Finally, a systematic analysis of the image resolution (estimated by Fourier ring correlation) as a function of the number of acquired frames (and therefore the total effective exposure time) reveals that the diffraction-limited \SI{5.2}{\nm} resolution is achieved at around 5 minutes of total exposure time, a value that will eventually be reduced in the future with the advent of more brilliant and coherent fourth generation synchrotron sources.
It must be noted that while the resolution reaches the diffraction limited value of \SI{5.2}{\nm} after only 5 minutes of exposure, increasing the exposure time beyond this value still increases the signal to noise ratio of the reconstructions, as can be observed in FRC curves [Fig.~\ref{fig:figS5}(a)] or in reconstructions at different exposure times [Fig.~\ref{fig:figS6}(a)].

\subsection{Micromagnetic Simulations}
\begin{figure}[h]
\centering
\includegraphics[width= 13 cm]{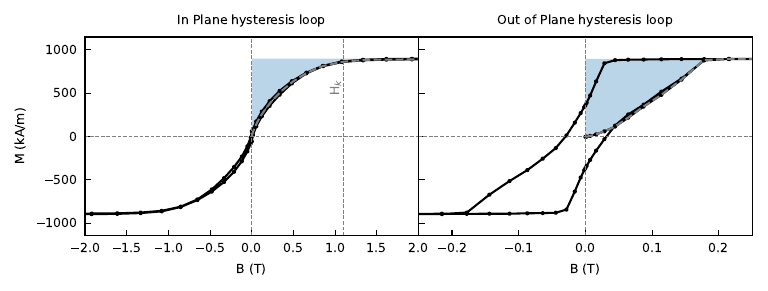}
\caption{Magnetic parameter measurement. Hysteresis loops of the multilayer acquired with SQUID. Light-coloured areas display the integrals used to compute the material's magnetic anisotropy using the method employed in \cite{PhysRevB.87.134422}.}
\label{fig:figS7}
\end{figure}

The micromagnetic simulations were performed using the solver Mumax3 (version 3.10)~\cite{Vansteenkiste2014}. The magnetic parameters were taken from previous works on similar systems (exchange parameter $A=\SI{11}{\pico\joule\metre^{-1}}$ \cite{ajejas_interfacial_2022}, DMI 
$D=\SI{1}{\milli\joule\metre^{-2}}$ \cite{legrand_room-temperature_2017}) or obtained from SQUID measurements (saturation magnetization $M_s=\SI{893}{\kilo\ampere\metre^{-1}}$, uniaxial anisotropy $K_u=\SI{752}{\kilo\joule\metre^{-3}}$, see Figure~\ref{fig:figS7}).

\begin{figure}[h]
\centering
\includegraphics[width=7cm]{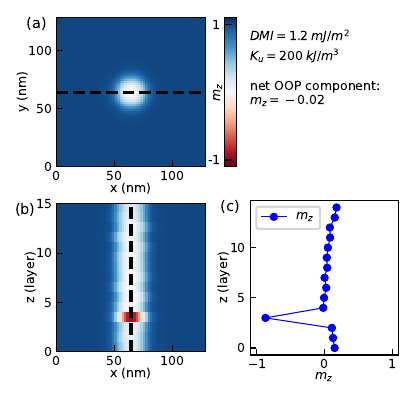}
\caption{Determination of relevant quantities from micromagnetic simulations. (a) $m_z$ averaged over the z axis a for defects with DMI $=\SI{1.2}{\milli\joule\metre^{-2}}$ and $K_u=\SI{200}{\kilo\joule\metre^{-3}}$. (b) Vertical section of $m_z$ along the dashed black line in (a). (c) Linescan of $m_z$ along the dashed black line in (b), at the center of a defect.}
\label{fig:figS8}
\end{figure}

We modeled a slab with 64\texttimes64\texttimes45 cells of sizes \SI{2}{\nm}\texttimes\SI{2}{\nm}\texttimes\SI{1.1}{\nm}. To take into account the presence of a non-magnetic buffer between the Co layers (\SI{2}{\nm} of Pt and \SI{0.5}{\nm} of Cu), we divided each trilayer into three layers, one magnetic and two non-magnetic ones. The magnetic parameters were rescaled using the effective medium model \cite{lemesh_accurate_2017} as in the following. A scaling factor $f=t_{Co}/p=0.73$ was defined, where $t_{Co}$ and $p$ are the thickness of the magnetic layers and of the simulation layers, respectively. Then the saturation magnetization, the DMI and exchange stiffness were multiplied by $f$, while the anisotropy was rescaled following the formula $K_u=K_u \cdot f - \mu_0 M_s^2(f-f^2)$.
The localized variations in magnetic properties were modeled by inserting a cylindrical region of \SI{30}{\nm} in radius having different magnetic parameters (DMI and $K_u$) with respect to the material's intrinsic properties.
The magnetic texture was relaxed from a random configuration to obtain a demagnetized state. Then, the external magnetic field was ramped up to \SI{210}{\milli\tesla} to verify the presence of defect-localized magnetic textures at high fields. The net out-of-plane component of the magnetization $m_{z}$ at the center of the region was measured in order to obtain the phase diagram shown in Figure~4(b), which sums the results of 10 different simulations. The process is illustrated in Figure~\ref{fig:figS8} for one combination of defect-associated magnetic parameters $K_u$ and DMI.

To estimate the domain wall width parameter \textdelta\ at different $K_u$ and DMI, the anisotropy and DMI were changed uniformly in the entire slab and the system was relaxed from a single domain wall state at remanence. \textdelta\ was measured by fitting the domain wall linescan as in the experimental images.

\end{appendix}

\begin{acknowledgments}
Financial support from the Leibniz Association via Grant No.~K162/2018 (OptiSPIN) and the Helmholtz Young Investigator Group Program (Grant VH-NG-1520) is acknowledged.
We acknowledge DESY. This research was carried out at PETRA III at the P04 beamline.
\end{acknowledgments}

\bibliography{references}
\onecolumngrid

\begin{table}[htbp]
\centering
\caption{\bf Relevant experimental parameters for the images showed in the paper}
\begin{tabular}{cccccccc}
\hline
 Figure & \ref{fig:fig3}(a,b) & \ref{fig:fig3}(c) & \ref{fig:fig3}(d)   & \ref{fig:fig3}(e) & \ref{fig:fig1}, \ref{fig:fig2}, \ref{fig:fig3}(f,g)\\
\hline
CCD distance (\SI{}{\cm}) & 15.8 & 15.8 & 15.8 & 26 & 8.4\\
CCD pixel size (\SI{}{\um}) & 20 & 20 & 20 & 13.5 & 13.5 \\
CCD pixel number  & 1340\texttimes1300  & 1340\texttimes1300 & 1340\texttimes1300   & 2048\texttimes2048 & 2048\texttimes2048\\
image pixel size (\SI{}{\nm})  & 9.9  & 9.9 & 9.9 & 11 & 5.3\\
total exposure time (\SI{}{\s})   & 4 & 1706  & 1022 & 756 & 2446 \\
\hline
\end{tabular}
  \label{tab:images_exp_details}
\end{table}

\begin{table}[htbp]
\centering
\caption{\bf Beam-stops used and associated data}
\begin{tabular}{ccc}
\hline
\multicolumn{3}{c}{Fig.~\ref{fig:fig1},\ref{fig:fig2},\ref{fig:fig3}(f,g)} \\
\hline
 & small beam-stop & large beam-stop \\
\hline
beam-stop diameter (\SI{}{\mm}) & $\sim$0.6 & $\sim$2\\
number of frames (per polarization) & 414 & 187 \\
exposure time per frame (\SI{}{\s}) & 0.7-1.1 & 4.7 \\
total exposure time per polarization (\SI{}{\s}) & 342 & 881 \\
\hline
\end{tabular}
\label{tab:images_exp_times}
\end{table}

\end{document}